\documentclass[a4paper]{jpconf}
\def\ut#1{\mathop{\vtop{\ialign{##\crcr
     $\hfil\displaystyle{#1}\hfil$\crcr\noalign
     {\kern1pt\nointerlineskip}\hbox{$\hfil\sim\hfil$}\crcr
     \noalign{\kern1pt}}}}}

\def\undersymbol#1#2{\mathop{\vtop{\ialign{##\crcr
     $\hfil\displaystyle{#2}\hfil$\crcr\noalign
     {\kern1pt\nointerlineskip}\hbox{$\hfil#1\hfil$}\crcr
     \noalign{\kern1pt}}}}}

\def\degr{^0}

\usepackage{graphicx}
\begin{document}
\title{CMB as a possible new tool to study the dark baryons in galaxies}

\author{F. De Paolis, G. Ingrosso, A.A. Nucita, D. Vetrugno}
\address{Dipartimento di Fisica, Universit\`a del Salento, Via per Arnesano, I-73100, Lecce, Italy and\\
INFN, Sezione di Lecce, Via per Arnesano, I-73100, Lecce, Italy}
\ead{francesco.depaolis@le.infn.it}

\author{V.G. Gurzadyan, A.L. Kashin, H.G. Khachatryan, S. Mirzoyan}
\address{Yerevan Physics Institute and Yerevan State University,
Yerevan, Armenia}

\author{Ph. Jetzer}
\address{Institute f\"ur Theoretische Physik, Universit\"at Z\"urich, Winterthurerstrasse
190, 8057 Z\"urich, Switzerland}

\author{A. Qadir}
\address{Center for Advanced Mathematics and Physics, National
University of Sciences and Technology, H-12, Islamabad, Pakistan}

\begin{abstract}
Baryons constitute about $4\%$ of our universe, but most of them
are missing and we do not know where and in what form they are
hidden. This constitute the so-called missing baryon problem. A
possibility is that part of these baryons are hidden in galactic
halos. We show how the 7-year data obtained by the WMAP satellite
may be used to trace the halo of the nearby giant spiral galaxy
M31. We detect a temperature asymmetry  in the M31 halo along the
rotation direction up to about 120 kpc. This could be the first
detection of a galactic halo in microwaves and may open a new way
to probe hidden baryons in these relatively less studied galactic
objects using high accuracy CMB measurements.
\end{abstract}

\section{Introduction}

We have entered in a new era of high precision cosmology and it is
clear by primordial nucleosynthesis constraints and by the results
from the power spectrum of the CMB produced by the 7-years data of
WMAP satellite that our universe is composed of $\simeq 73\%$ of
dark energy (DE), $\simeq 23\%$ of non-baryonic dark matter (DM)
and for only $\simeq 4\%$ by baryons. About baryons, we know that
about $10\%$ of them are in stars (i.e. in the visible part of
galaxies); the hot gas in galaxies and galaxy clusters accounts
for another $20-30\%$ of baryons; but about $60-70\%$ of all
baryons are missing and is not known where and in which form they
are hidden \footnote{Observations also show that galaxies are
missing most of their baryons with respect to the cosmological
baryon/matter ratio (see e.g.
\cite{corbelli2003,hoekstra,bregman2007}).}. A possibility is that
these baryons are contained in the so-called cosmic filaments in
the form of a warm-hot intergalactic medium (see e.g.
\cite{cen99,cen2006,McKelvie,nicastro}). However, it is unlikely
that all the missing baryons are confined in the cosmic filaments
and, actually, there are many reasons to believe that a
non-negligible amount of these hidden baryons are in galactic
halos (see below in this section) or even in the galactic disk
(see e.g. \cite{revaz} and references therein) \footnote{The first
to suggest the concept of the presence of the gas above the disk
of the Milky Way was Spitzer in his remarkable paper of 1956
\cite{spitzer} where he also discussed the nature and the possible
evolution of the gas eventually present there.}.

A rather clear piece of evidence for the existence of large
amounts of gas around galaxies is the widespread detection of
absorption lines in front of quasars, the so-called Ly-$\alpha$
forest, that show hundreds of absorption systems along a single
line of sight. Also damped Ly-$\alpha$ systems with hydrogen
columns density larger than about $10^{20}$ cm$^{-2}$ are
associated with galaxies and both systems imply that galaxies at
high redshift ($\sim 2-2.5$) would have huge gas halos around
them. Where is all that gas now? One possibility is that it has
been expelled from galaxies during the first chaotic phase of star
formation into the cosmic filaments. However, that might have
happened only in part, leaving nowadays some of that gas around
galactic disks.

There is further evidence that there is some amount of gas in
galactic halos, and in particular in that of our galaxy: the
problem is to measure how much it is. That is, obviously,  not so
easy. Let us briefly discuss some of this evidence.

Galaxies seem to evolve along the Hubble sequence from Sc type,
characterized by small bulges and open spiral arms to Sa,
characterized by larger bulges and tighter spirals. During this
evolution the $M/L$ ratio, which is a measure of the amount of
dark matter, decreases (while the number of stars increases)
implying that part of the dark matter should transform into stars
and that means that part of the dark matter should consist of
baryons in gaseous form. Moreover, star-forming disk galaxies
similar to our Milky Way, should exhaust their cold gas reservoir
(necessary to make stars) within a few Gyrs unless it is
replenished in some way \cite{chiappini,sommer-larsen}. Numerical
simulations (see e.g. \cite{twarog})  indeed show that galaxies
like the Milky Way  convert about $1~M_{\odot}$ yr$^{-1}$  of gas
into stars while their gas content remains approximately
unchanged. The reservoir of the accretion material is as yet
unidentified but it goes without saying that it should come from
the region  that surrounds the galactic disks and might even
reside in the intergalactic medium (see \cite{binney2009} for a
discussion about the cooling mechanisms of the hot intergalactic
gas). In any case, a gas accretion rate of about $1~M_{\odot}$
yr$^{-1}$ for M31/Milky Way galaxies onto their galactic disks is
compulsory (see also \cite{richter2011}).

The wealth of data in the last two decades show that there is good
evidence for extra-planar gas around the Milky Way and other
galaxies  and it is detected in all gaseous phases: neutral, warm
atomic, molecular and hot X-ray emitting gas. Since about a decade
we know that around the Milky Way and Andromeda (M31) galaxies
there is a population of high-velocity clouds (HVCs) that extends
up to $10-20$ kpc and are not seen beyond  about $50$ kpc
\cite{wakker2008}. There are also intermediate-velocity clouds
(IVCs) with a vertical scale height of $4-5$ kpc and a total mass
about $3-4\times 10^8~M_{\odot}$, forming a population separated
with respect to that of the HVCs. The innermost halo clouds show a
metallicity that is about half of the solar one and this favors
the Galactic fountain model for its origin in which hot gas is
ejected out of the galactic disk, cools down and falls back onto
the disk (see e.g. \cite{richter}). As far as HVCs are concerned,
since generally only the radial velocity can be estimated for
these objects and distance information are sparse and not well
known, a closed theory about their origin is not available today.
One possibility is that they represent metal poor gas from the
intergalactic space falling onto the galactic disk \cite{oort},
the other is that they represent the remnant of the galactic halo
formation.

Also around the Andromeda galaxy (M31) deep radio synthesis
observations has shown the presence of gas clouds within about 50
kpc. These clouds, which are the analogue of the HVCs, have size
of about 1 kpc and mass $\simeq 10^5~M_{\odot}$ \cite{westmeier}.
The total mass in HVCs is about $10^8~M_{\odot}$
\cite{richter2011}.

It also seems that a dilute gas halo at a temperature $\sim 10^6$
K is present around some spiral galaxies. Elliptical galaxies (and
particularly bright ellipticals), as is well known, do contain
large amount of hot diffuse gas emitting in X-rays (see e.g.
\cite{dis} and references therein). For spirals, evidence is much
less convincing. A difficulty in this respect is that the emission
measure scales with the square of the electron density and once
the galaxy surface brightness drops to the level of the X-ray
background (mainly due to the Milky Way) further detection is
impossible. That is the reason why the size of the X-ray emitting
gas around spiral galaxies is very poorly constrained.
Interestingly enough, very recently a hot gaseous halo has been
detected by XMM-Newton satellite towards UGC12591, the fastest
rotating spiral galaxy \cite{dai2011}, up to a distance of about
110 kpc from the galaxy center. Combining the X-ray data with
near-IR and radio measurements it has been found that the baryon
mass fraction in this galaxy is about $3-4\%$ (in particular the
baryon mass within about 500 kpc is $\simeq 5.9 \times
10^{11}~M_{\odot}$ while the total mass, estimated using the
 galaxy rotation curve, is found to be $\simeq 2.7 \times
10^{13}~M_{\odot}$). This would imply that the majority of the
missing baryons in spiral galaxies does not reside in their hot
gas halos \footnote{The fact that galactic halos do not contain
much hot baryons is also confirmed by the dispersion measure of
pulsars in the LMC \cite{anderson}.}.

Indeed, after the detection of the first microlensing events
towards the LMC \cite{eros,macho} \footnote{Microlensing
observations conducted since two decades both towards the Galactic
bulge and the Magellanic Clouds by MACHO, EROS and OGLE
Collaborations have shown that a (probably) not-negligible
fraction (about $5-20\%$) of the halo dark matter may be made by
MACHOs (see e.g. \cite{alcock2000,tisserand,calchinovati2007})
although this result as well as the estimation of the average
MACHO mass implied by observations depends on the specific adopted
halo model (see e.g. \cite{dij,grenacher}). Pixel-lensing
observations towards the M31 galaxy have been conducted by several
collaborations leading to the discovery of more than 35 events
\cite{ansari1999,auriere2001,novati2002,riffeser2003,dejong2004,novati2007,novati2009}
up to now (one of which  even showed the presence of an exoplanet
in Andromeda galaxy \cite{ingrosso2009,ingrosso2011}) give
uncertain conclusions about the fraction of the halo dark matter
in the form of MACHOs, ranging from about $20\%$ for
$0.1-1~M_{\odot}$ MACHOs to the possibility of explaining the
detected events simply by self-lensing (see e.g.
\cite{novati2005,depaolis2005,dejong2006,ingrosso2006,ingrosso2007,novati2010}).
} a model was proposed for the formation of MACHOs (Massive
Astrophysical Compact Halo Objects) in the galactic halo
\cite{paperdijr95a,paperdijr95b}. Actually, this model  naturally
emerges from the present-day understanding of the globular cluster
formation. Indeed, the Fall-Rees theory for the formation of
globular clusters \cite{fall} predicts, without any further
assumption, that dark clusters made of brown dwarfs and cold gas
clouds should lurk in the galactic halo at galactocentric
distances larger that $10 - 20$ kpc. Accordingly, the inner halo
is populated mainly by globular clusters, whereas the outer
galactic  halo should be dominated by dark clusters populated by
MACHOs and cold gas clouds \footnote{Somewhat similar scenarios
have been also proposed in \cite{ashman,gs,draine,walker}. Models
in which the galactic dark matter is in baryonic form and
distributed in a thin disk have also been proposed along with many
observational consequences \cite{pfenniger,sciama}.}. The gas
clouds in the dark clusters should be very cold, with a
temperature very close to that of the CMB, should have sub-solar
mass  and size about that of the Solar System (for details see
\cite{apj98}).

A novel way of investigating the amount and distribution of the
dark baryons in the galactic halos, and in particular in that
around the M31 galaxy is discussed  in the next section. This
method deals with the use of CMB data (at present the 7-year WMAP
data) to trace galactic disks and halos.

\section{The temperature  asymmetry in the M31 halo by the 7-year WMAP data}

Galactic disks are well studied objects in all wavelengths and
give important information on the mass distribution within and
around galaxies \cite{binney}. On the other hand, galactic halos
are relatively less studied structures and there are still many
ambiguities not only in the main halo constituents, but also in
the basic properties such as, in particular, the rotation.

The degree to which galactic halos rotates with respect to the
disks is a relevant and difficult issue to be investigated. The
rotation of the galactic halos, indeed, is clearly related to the
formation scenario of galaxies. In the standard collapse model
(see e.g. \cite{eggen}) both the halo and disk of galaxies derive
from the same population. The rotation  of the outer halo should
be, in this case, aligned with the disk angular momentum. On the
contrary, in a hierarchical formation scenario, structures
arriving later in the outer halo should have  a minor connection
to the disk. Therefore, it is evident that information on the
galactic halo rotation provide key insights about the formation
history  of galaxies. Nevertheless, it is important to stress that
testing for the rotation even of the closest galaxy (M31) halo is
still beyond our reach \cite{courteau} (see also \cite{deason}).

The first attempt of using the 7-year WMAP data \cite{J} in the
three bands W (94 GHz),  V (61 GHz), and Q (41 GHz) to map in
microwaves both the disk and halo of the M31 galaxy is provided in
\cite{depaolis2011} \footnote{The use of three WMAP bands is
important in revealing the possible contribution of the Galactic
foregrounds since dust, free-free, and synchrotron emission
contributes in a different way in each band. We also remind that
the band least contaminated by the synchrotron radiation of the
Galaxy is the W-band, which also has the highest angular
resolution.}.
To reveal the different contributions by the M31 disk and halo,
the region of the sky towards the M31 galaxy has been divided into
several concentric circular areas as shown in Fig. 1 in
\cite{depaolis2011}.

The M31 disk does contain gas observed mainly at 21 cm wavelength
(but also in the IR) up to a distance of about 40 kpc from the
galactic center. It is also well known that the M31 disk rotate
with a speed of about 250 km s$^{-1}$ and this has been clearly
shown also by the velocity maps provided in
\cite{chemin,corbelli}. In \cite{depaolis2011} a temperature
asymmetry along the direction of the M31 rotation has been
observed for the first time also in microwaves with a maximum of
$\simeq 130~\mu$k/pixel at about 20 kpc from the M31 center. This
temperature asymmetry is very likely induced by the Doppler shift
effect due to the M31 disk rotation speed. The robustness of this
result has been tested by considering 500 randomly distributed
control fields in the three WMAP bands and also by simulating 500
sky maps from the best fit cosmological parameters. Both
procedures give comparable results and imply that there is less
than $\simeq 2\%$ probability that the signal is due to a random
fluctuation of the CMB signal. An analogous study has been also
conducted in the same paper towards the M31 halo within $20\degr$
(about 240 kpc from the M31 center). We found also towards the M31
halo  a CMB temperature asymmetry up to about 120 kpc with a peak
temperature contrast of about $40~\mu$K/pixel. Although the
confidence level of the signal, if estimated purely statistically
(i.e. with 500 control fields and 500 simulated sky maps), is not
high (we find, indeed, that there is a probability of less than
about $30\%$ that the detected temperature asymmetry in the M31
halo is due to a random fluctuation of the CMB signal) \footnote{
Actually, if one takes the direction of rotation of the M31 disk
into account, such a probability reduces (by using the theorem of
the composite probability) by a factor of two. So that one can
conclude that there is a probability of less than about $15\%$
that the detected temperature asymmetry in the M31 halo is due to
a random fluctuation of the CMB signal.}, the geometrical
structure of the temperature asymmetry in the three bands point
towards a real effect modulated by the rotation of the M31 halo.

We point out once more that the use of three WMAP bands is
important for revealing the role of the contribution of the
Galactic foregrounds since each emission mechanism contributes
differently in each band. The fact that the temperature contrast
seems present in all three bands and is more or less the same in
each band up to about $10\degr-11\degr$ (about 120 kpc from the
M31 center) indicates that the foregrounds are far weaker than the
effect. A size of about 120 kpc corresponds to the typical size
inferred for the dark matter halos around massive galaxies and
might open the possibility of a new way of studying these systems,
both galactic disks and halos, at microwave wavelengths. In any
case, a careful analysis of the Planck data that should be
released shortly should allow either to prove or disprove our main
results.

\section{Discussion}

As clear from the discussion above, the detected temperature
asymmetry for the M31 disk is fairly clear in all WMAP bands, and
is also expected due to the foreground emission of the M31 disk
modulated by the Doppler shift induced by the disk rotation.
Incidentally, the M31 galaxy has been recently detected by the
Planck observatory \cite{planckm31} \footnote{However, there is no
mention of any temperature asymmetry in the M31 disk in that
paper.}, whereas it did not appear in the WMAP list. These are all
reasons to expect that the particular effect we discuss here can
be studied more accurately with Planck data.

As for the M31 halo, we have shown that, although less evident
than for the M31 disk, there is less than about $20\%$ probability
that the detected temperature asymmetry at a galactocentric
distance $\sim 50$ kpc comes from a random fluctuation of the CMB
signal. \footnote{We  also mention that the number and the
temperature profile of radio sources in CMB maps \cite{gurzadyan}
excludes their significant contribution in the effect under
study.}

If one assumes that this temperature asymmetry in the M31 halo
relies in the M31 itself and is related to the M31 halo rotation,
a natural question that arises is about the origin of this effect.
In all generality, four possibilities may be considered: ($i$)
free-free emission; ($ii$) synchrotron emission; ($iii$)
Sunyaev-Zel'dovich (SZ) effect; and ($iv$) cold gas clouds
populating the M31 halo. The first three effects assume the
presence of a rather hot plasma in the halo of M31. Although this
hot plasma has not been detected yet, one can assume that a
certain amount of this plasma  can populate the M31 halo (spiral
galaxies are believed to have much less hot gas than ellipticals)
and may rotate with a certain speed. Free-free emission arises
from electron-ion scattering while synchrotron emission comes
mostly from the acceleration of cosmic-ray electrons in magnetic
fields. Both effects give rise to a thermal emission with a rather
steep dependence on the frequency \cite{bennett} that therefore
should give a rather different temperature contrast in the three
WMAP bands. The absence of this effect indicates that the
contribution from possibilities ($i$) and ($ii$) should be
negligible. In the case of ($iii$), even for typical galaxy
clusters with diffuse gas much hotter than that possibly expected
in the M31 halo, the rotational effect produces a temperature
asymmetry of at most a few $\mu$K pixel$^{-1}$, depending on the
rotational velocity and the inclination angle of the rotation axis
\cite{cooray}, that is much less than that observed towards the
M31 halo. Actually, a possible temperature asymmetry in the CMB
data towards the M31 halo as a consequence of the existence of a
population of cold gas clouds in its halo was predicted in
\cite{paperdijqr95c} - possibility ($iv$). Indeed, if the halo of
the M31 galaxy would contain gas clouds, one expects them to
rotate along the disk rotation (even if, perhaps, more slowly),
and thus there should be a Doppler shift inducing a temperature
anisotropy $\Delta T$ between one side of the M31 halo and the
other with respect to the rotation axis perpendicular to the disk.
In the case of optically thin halo clouds the Doppler induced
temperature asymmetry would be ${\Delta T}/{T_r} \simeq 2{v}S\bar
\tau /c~$, where $T_r\simeq 2.725$ K is the CMB temperature, $v$
is the M31 averaged rotation speed, $\bar \tau$ the averaged cloud
optical depth over the frequency range ($\nu_1 \leq \nu \leq
\nu_2$) of a certain detection band, and $S$ the cloud filling
factor, i.e. the ratio of filled (by clouds) to total projected
sky surface in a given field of view. In the case of optically
thick gas clouds instead one has ${\Delta T}/{T_c} \simeq
2{v}S/c$, where $T_c$ is the cloud temperature \footnote{Note,
however, that cold gas clouds in galactic halos are not expected
to be optically thick at microwave wavelengths.}.

In \cite{paperdijqr95c} we concentrated mainly on the optically
thick option and, as a suggested test for the proposed model (see
also \cite{depaolis1996,depaolisijmpd}), made a rough estimate
\footnote{We estimated, based on our model, S to be about 1/25 and
assumed a halo rotation speed about 100 km s$^{-1}$. By the way,
this value for S is not far from that coming from recent
observations of HVCs in M31 presented in Fig. 1 in
\cite{richter2011}. Indeed it turns out that S can be
parameterized as $S\simeq 2.1{\rm exp}[-r/(12 kpc)]$, so that
$S\simeq 0.05$ at about 40 kpc from the M31 center. We emphasize,
however, that this limit has to be considered as a lower limit to
the true value of S since it only comes from considering the
observed HVCs in the M31 halo.} of the expected temperature
asymmetry between the two opposite sides of the M31 halo as
${\Delta T}/{T_r} \simeq 2\times 10^{-5}\bar \tau$ that implies a
temperature asymmetry about $50~\mu$K for $\bar \tau \ut < 1$,
that is very close to the value recently found in
\cite{depaolis2011}.

Before closing this Section we would like to mention that one of
the most serious problems associated with the presence of cold gas
clouds  in galactic halos is that of explaining their mere
dynamical stability. As a matter of fact, isothermal perfect gas
clouds without fixed boundaries are unstable with respect to the
gravothermal catastrophe on time scales of a few crossing times.
Generally speaking, the outcome of the gravitational collapse of a
gas cloud is the formation of a dense central core (see e.g.
\cite{larson1969}). At the beginning, the evolution is almost
isothermal (due to the low optical depth of the gas cloud) but,
when opacity exceeds unity, the central temperature and pressure
rapidly increase, leading to star formation. However, halo cold
gas clouds may mostly succeed in avoiding collapse since in low
temperature and low opacity conditions, collapse may lead to the
formation of solid/liquid $H_2$ (below about 14 K) and to the
formation of a temperature inverted gas cloud with a condensed
cold core. This new physical ingredient provides the possibility
of stabilizing these cold gas clouds. Such loosely bound cold
clouds could form a gas reservoir in galaxy halos and behaves as a
collisionless ensemble of matter. Incidentally, these cold cores
might also evaporate due to ambient heating, but they can leave
for Gyrs in the clod and low excitation environment such as those
expected in galactic halos (for further details see also
\cite{pfenniger1994,pfenniger2004}). Interestingly enough, cosmic
rays and energetic photons (for example those in the UV band of
the electromagnetic spectrum) might cause ionization in the solid
molecular hydrogen leading to the formation of $H_6^+$ and
$(HD)_3^+$. Very recently, it has been shown \cite{lin2011} that
the so-called unidentified infrared (UIR) bands, usually thought
to be related to  large molecules such as polycyclic aromatic
hydrocarbons, may instead be produced by $(HD)_3^+$ and $H_6^+$ in
solid form, thus showing that solid $H_2$ may indeed be abundant
in quiet astrophysical environments and opening a new way of
searching for cold gas clouds.

\section{Concluding remarks}

As mentioned earlier, it is likely that there are no grounds to
assume that a unique solution may solve the hidden baryon problem.
It is instead plausible that it is a composite problem reflecting
the diversity of astrophysical conditions.

For example, globular clusters are mostly composed of baryons and
no evidence for  non-baryonic  mass is present in these objects.
Thus, baryons are well separated from non-baryonic dark matter in
the formation process of these structures. On the galaxy cluster
scale most baryons are in the form of a hot X-ray emitting gas
while in the inter-galactic medium most baryons might be in the
form of a warm-hot gas.

The wealth of data especially in the last decade shows that there
is good evidence for the presence in the halos of spiral galaxies
of gas in all gaseous phases: neutral, warm atomic, and hot X-ray
emitting gas  \cite{bregman1}. Atomic gas (often identified as
HVCs) is observed in the radio band (particularly at 21 cm) and
through absorption lines towards field stars and quasars. The hot
gas may be detected in X-rays but, unfortunately, searches for
cold gas clouds in galactic halos are more problematic. Various
attempts have been undertaken to detect such clouds as searching
for the presence of a gamma-ray halo \cite{dixon,depaolis99},
stellar scintillations \cite{moniez03,habibi}, obscuration events
towards the LMC \cite{drakecook}, ortho-$H_2D^+$ line at 372 GHz
\cite{ceccarelli}, and extreme scattering events in quasar
radio-flux variations \cite{walker}. All these searches have given
no clear indication of cold gas cloud presence in galactic halos,
but these searches are going on using more and more sensitive
facilities and new developments, such as searching for the
vibrational transition lines produced by $(HD)_3^+$ and $H_6^+$ in
solid form. Therefore, the issue of a realistic estimate of the
amount of baryonic mass in galactic halos and their size does
remain open and a new perspective in this research can be given by
investigating it in microwaves following \cite{depaolis2011}.

\ack{VG and AQ acknowledge some financial support by the
Department of Physics, University of Salento, and by INFN.}

\end{document}